\documentclass[12pt,a4]{article}

\usepackage{axodraw}
\usepackage{psfig}
\usepackage{latexsym}
\usepackage[dvips]{graphicx}
\input{epsf}

\newcommand{\be}{\begin{equation}}
\newcommand{\ee}{\end{equation}}
\newcommand{\bea}{\begin{eqnarray}}
\newcommand{\eea}{\end{eqnarray}}

\title{\Large \bf A 2 Loop 2PPI Analysis of $\lambda \phi^4$ at Finite
Temperature}
\author{\normalsize G. Smet, T. Vanzieleghem, K. Van Acoleyen, H. 
Verschelde\\
\it University of Gent\\
\it Department of Mathematical Physics and Astronomy \\
\it Krijgslaan 281-S9 \\
\it B-9000 Gent, Belgium}
\date{ }
\begin{document}
\maketitle
\pagenumbering{arabic}
\vskip 36pt
\begin{abstract}

We calculate the finite temperature effective potential of $\lambda \phi^4$ at the two
loop order of the 2PPI expansion. This expansion contains all diagrams which remain connected
when two lines meeting at the same point are cut and therefore sums systematically the
bubble graphs. At one loop in the 2PPI expansion, the symmetry restoring phase transition 
is first order. At two loops, we find a second order phase transition with mean field
critical exponents.

\end{abstract}
\newpage

\section{Introduction}

The $\lambda \phi^4$ theory at finite temperature is of great
interest for the field of phase transitions in the early
universe and heavy ion collisions. When used as a simple model
for the Higgs particle in the standard model of electroweak
interactions, it may allow the study of symmetry breaking phase
transitions in the early universe. For N=4 scalar fields, it is
also a model of chiral symmetry breaking in Q.C.D. and hence
relevant for the theoretical study of heavy ion collisions.
Moreover, this theory is an excellent theoretical laboratory
where analytic non-perturbative methods can be tested.

We recently [1,2] introduced the 2PPI expansion as an analytic
perturbative variational method in quantum field theory. It is
perturbative in the sense that one calculates Feynman diagrams
which remain connected when two internal lines meeting at the
same point are cut (2PPI = two particle point irreducible). It
is variational because the contribution of 2PPR diagrams is
absorbed in a variational effective mass. It can be compared to
the CJT [3] formalism which is an expansion in 2PI diagrams
where the variational object is the two particle Greens
function. The amount of resummation in the 2PPI expansion is
less than in the 2PI expansion but because the variational
object is just a mass parameter, it has the advantage of
being much more tractable at higher order. The 2PPI expansion
has also some other advantages when compared with other
resummation methods. Besides being a systematic expansion, its
renormalization is straightforward. In [2], it was shown that
the 2PPI expansion can be renormalized with the usual
counterterms in a mass independent scheme. The simplest approach
is to calculate the 2PPI diagrams using dimensional
regularization and subtract the pole terms. The 2PPI expansion
is especially well suited for finite temperature field theory.
It does not have the well known problems with perturbation
theory at high temperature because the contribution of hard
thermal loops can be absorbed in the variational effective mass.
Also, the Goldstone theorem is obeyed at any order of the
expansion [4].

In a recent paper [4], we applied the 2PPI expansion to the
calculation of the effective potential of $\lambda \phi^4$
theory at finite T. At one loop 2PPI, we showed equivalence with
summation of daisy and superdaisy diagrams [5,6,7]. As is well
known, at this order of resummation, the phase transition is
first order. In this paper, we want to do the full two loop
calculation of the effective potential in the 2PPI expansion.
The only two loop 2PPI diagram which we have to calculate is the
setting sun diagram. As it turns out, this two loop contribution
is large enough to turn the first order transition into a second
order one as expected from lattice simulations and experiment.
The organisation of this paper is as follows. In
section 2, we give a short introduction to the 2PPI expansion
and its renormalization. In section 3, we calculate the one and
two loop terms of the 2PPI expansion of the effective potential.
In section 4, we discuss our numerical results and show that the
phase transition is second order with mean field critical exponents.

\section{The 2PPI expansion} 

Let us consider a generic 2PPR diagram (two particle point
reducible), i.e. a diagram that becomes disconnected when two
internal lines meeting at the same point are cut (see fig. 1).
It is clear that the 2PPR insertions are seagull (2 external
$\phi$ lines) or bubble graphs. From that, we can conclude that
the 2PPI expansion sums seagull and bubble graphs. How is this
summation carried out ? We first notice that the seagull and
bubble graphs contribute to the selfenergy as effective masses
$\frac{\lambda}{2} \langle \phi \rangle^2$ and
$\frac{\lambda}{2} \langle \phi^2 \rangle_c = \frac{\lambda}{2}
\Delta$ respectively. Therefore, one could naively think that
all 2PPR insertions are summed by simply deleting the 2PPR
graphs from the 1PI expansion and introducing the effective mass,
\be
\overline{m}^2 = m^2 + \frac{\lambda}{2} (\varphi^2 + \Delta)
\ee
where $\varphi = \langle \phi \rangle$, in the remaining 2PPI
graphs. This is not correct since there is a double counting
problem which can be easily understood in the simple case of the
2loop vacuum graph (daisy graph with two petals) of fig. 2.a.
Each petal can be seen as a self energy insertion in the other,
so there is no way of distinguishing one or the other as the
remaining 2PPI part. We could however earmark one of the petals
by applying a derivative with respect to $\varphi$ (fig. 2.b).
This way the 2PPI remainder (which contains the earmark) is
uniquely fixed. Now, there are two ways in which the derivative
can hit a $\varphi$ field. It can hit an explicit $\varphi$
field which is not a wing of a seagull or it can hit a wing of a
seagull, i.e. an implicit $\varphi$ field hidden in the
effective mass. We therefore have
\be
\frac{\delta}{\delta \varphi} \Gamma^{1PI}_q (m^2,\varphi) = \frac{
\partial}{\partial \varphi} \Gamma^{2PPI}_q(\overline{m}^2,\varphi) +
\lambda \varphi \frac{\partial}{\partial \overline{m}^2}
\Gamma^{2PPI}_q (\overline{m}^2,\varphi)
\ee
where $\Gamma^{1PI} = S(\varphi) + \Gamma^{1PI}_q$. Using the
definition (1) of the effective mass, we can rewrite this as :
\be
\frac{\delta}{\delta \varphi} \Gamma^{1PI}_q (m^2,\varphi) =
\frac{\delta}{\delta \varphi} \Gamma^{2PPI}_q
(\overline{m}^2,\varphi) - \frac{\lambda}{2} \frac{\delta
\Delta}{\delta \varphi} \frac{\partial \Gamma^{2PPI}_q}{\partial
\overline{m}^2} (\overline{m}^2,\varphi)
\ee
Using an analogous combinatorial argument we have
\be
\frac{\partial \Gamma^1_q}{\partial m^2} (m^2,\varphi) =
\frac{\partial \Gamma^{2PPI}_q}{\partial \overline{m}^2}
(\overline{m}^2,\varphi) 
\ee
and since $\Delta/2 = \frac{\partial \Gamma^{1PI}_q}{\partial
m^2}$, we find the following gap equation for $\Delta$ :
\be
\frac{\Delta}{2} = \frac{\partial \Gamma^{2PPI}_q}{\partial
\overline{m}^2} (\overline{m}^2,\varphi)
\ee
The gap equation (5) can be used to integrate (3) and we finally
obtain :
\be
\Gamma^{1PI}(m^2,\varphi) = S(\varphi) + \Gamma^{2PPI}_q
(\overline{m}^2, \varphi) - \frac{\lambda}{8} \int d^D x
\Delta^2 
\ee
with $\overline{m}^2$ the effective mass given by (1). So
indeed, the 2PPR insertions can be summed into the effective
mass $\overline{m}^2$ but there is a corection term (the last
term in (6)) which accounts for double counting. This effective
mass can be determined self consistently from the gap equation
(5) which can be rewritten as :
\be
\overline{m}^2 = m^2 + \frac{\lambda}{2} \varphi^2 + \lambda
\frac{\partial \Gamma^{2PPI}_q}{\partial \overline{m}^2}
(\overline{m}^2,\varphi) 
\ee

Up till now, we have remained silent about renormalization. For
example, $\Delta$ is quadratically divergent and needs
renormalization. The question is if after renormalization, the
simple relation (6) and the gap equation (5) remain valid.
Fortunately, this is so if one uses a mass independent
renormalization scheme. In [2] we showed that using dimensional
regularization (which eliminates quadratic divergences) and a
mass independent renormalization scheme, $\Delta$ can be
renormalized as :
\be
\Delta_R = Z_2 \Delta + \delta Z_2 \varphi^2 + \frac{2\delta
Z^{2PPR}_2}{\lambda} m^2
\ee
where renormalization constants come from the standard
counterterm Lagrangian for $\lambda \phi^4$ :
\be
\delta {\cal L} = \frac{\delta Z}{2} (\partial_{\mu} \phi)^2 +
\frac{\delta Z_2}{2} m^2 \phi^2 + \delta Z_{\lambda}
\frac{\lambda \phi^4}{4!}
\ee
and $\delta Z^{2PPR}_2$ contains only the divergences from mass
renormalization dia\-grams which are 2PPR. Using (8), one shows
that the gap equation (5) becomes after renormalization :
\be
\frac{\Delta_R}{2} = \frac{\partial
\Gamma^{2PPI}_{q,R}}{\partial \overline{m}^2_R}
(\overline{m}^2_R,\varphi) 
\ee
with $\overline{m}^2_R = m^2 + \frac{\lambda}{2}(\varphi^2 +
\Delta_R)$ and the relation (6) between $\Gamma^{1PI}$ and
$\Gamma^{2PPI}_q$ becomes
\be
\Gamma^{1PI}_R(\varphi,m^2) = S(\varphi) + \Gamma^{2PPI}_{q,R}
(\overline{m}^2_R, \varphi) - \frac{\lambda}{8} \int d^D x
\Delta^2_R 
\ee
The 2PPI effective action is renormalized solely with 2PPI
counterterms. The other 2PPR counterterms go into the
renormalization of the effective mass $\overline{m}^2$. 
In minimal subtraction, equation (8) is automatically fulfilled
and renormalization of the 2PPI expansion becomes extremely simple.

\section{Effective potential at two loop 2PPI}

From now on, we will work with the renormalized 2PPI expansion
and drop the subscript R, keeping in mind that modified minimal
subtraction ($\overline{MS}$ scheme) is implicity understood. The
first few diagrams in het 2PPI expansion are displayed in figure
3. We use the imaginary time formalism and the Saclay method [8]
to calculate the Matsubara sums. For the one loop contribution
to the 2PPI effective potential, we find :
\bea
V^{2PPI}_q(\overline{m}^2,\varphi) & = & \frac{1}{2} \sum \int
\ln (k^2 + \overline{m}^2) \nonumber \\
& = & \frac{\overline{m}^4}{64\pi^2} \left( \ln
\frac{\overline{m}^2}{\overline {\mu}^2} - \frac{3}{2} \right) +
Q_T (\overline{m}^2)
\eea
where $\sum \int = T 
\begin{array}[t]{c} \sum \\[-5pt]
\stackrel{n}{} \end{array} \int \frac{d^3k}{(2\pi)^3} ,
\overline{\mu}^2 = 4\pi e^{-\gamma} \mu^2$ and 
\be
Q_T(\overline{m}^2) = T \int \frac{d^3 q}{(2\pi)^3} \ln \left( 1 -
e^{-\frac{E_q}{T}} \right)
\ee
The effective potential at one loop 2PPI (sum of daisy and
superdaisy diagrams) is then given by :
\be
V_{eff}(m^2,\varphi,T) = \frac{m^2}{2} \varphi^2 +
\frac{\lambda}{4!} \varphi^4 + \frac{\overline{m}^4}{64\pi^2}
\left( \ln \frac{\overline{m}^2}{\overline{\mu}^2} -
\frac{3}{2}\right) + Q_T(\overline{m}^2) - \frac{\lambda}{8}
\Delta^2 
\ee
where $\overline{m}^2 = m^2 + \frac{\lambda}{2} (\varphi^2 +
\Delta)$ is a solution of the one loop gap equation
\be
\overline{m}^2 = m^2 + \frac{\lambda \varphi^2}{2} +
\frac{\lambda}{2} \left[ \frac{\overline{m}^2}{16\pi^2} \left(
\ln \frac{\overline{m}^2}{\overline{\mu}^2} - 1\right) + P_T
(\overline{m}^2) \right]
\ee
with
\be
P_T(\overline{m}^2) = 2 \frac{\partial}{\partial \overline{m}^2}
Q_T (\overline{m}^2) = \int \frac{d^3 q}{(2\pi)^3}
\frac{n_B(q)}{E_q} 
\ee

At two loops, the only 2PPI diagram we have to consider is the
setting sun diagram. It has been calculated together with the
other 2PPR two loop diagrams in [9]. For details, we refer to
the appendix, especially with respect to 2PPI renormalization.
For the effective potential at two loop 2PPI, we obtain the
following expression :
\bea
V_{eff}(m^2,\varphi,T) & = & \frac{m^2}{2} \varphi^2 +
\frac{\lambda}{4!} \varphi^4 + \frac{\overline{m}^4}{64\pi^2}
\left( \ln \frac{\overline{m}^2}{\overline{\mu}^2} -
\frac{3}{2}\right) + Q_T (m^2) \nonumber \\
& - & \frac{\lambda}{8} \Delta^2 - \frac{\lambda^2}{12} \varphi^2
(G_0(\overline{m}^2) +G_1(\overline{m}^2,T) +
G_2(\overline{m}^2,T)) 
\eea
where
\be
G_0(\overline{m}^2) = - \frac{3}{2} \frac{\overline{m}^2}{(4\pi)^4} \left(
\ln^2 \frac{\overline{\mu}^2}{\overline{m}^2} + 4 \ln
\frac{\overline{\mu}^2} {\overline{m}^2} + 4 + \frac{\pi^2}{6} -
\frac{8.966523919}{3} \right)
\ee
\bea
G_1(\overline{m}^2,T) & = & \frac{3}{(4\pi)^2} \left( \ln
\frac{\overline{\mu}^2}{\overline{m}^2} + 2 \right) \int
\frac{d^3 q_1}{(2\pi)^3} \frac{n_B(q_1)}{E_{q_1}} \nonumber \\
& + & \frac{3}{4(2\pi)^4} \int^{\infty}_0 dq_1 \frac{q_1
n_B(q_1)}{E_{q_1}} \int^{\infty}_0 \frac{dq_2}{E_{q_2}} \left( q_2 \ln
|\frac{X_+}{X_-}| - 4 q_1\right)\label{F1}
\eea
with 
\be
X_{\pm} = \left( E_{q_1} + E_{q_2} + E_{q_1\pm q_2}\right)^2
\times \left( - E_{q_1} + E_{q_2} + E_{q_1 \pm E_2}\right)^2 
\ee
\be
G_2(\overline{m}^2,T) = \frac{3}{4(2\pi)^4} \int^{\infty}_0 dq_1
\frac{q_1 n_B(q_1)}{E_{q_1}} \int^{\infty}_0 dq_2 \frac{q_2
n_B(q_2)}{E_{q_2}} \ln |\frac{Y_+}{Y_-}|\label{F2}
\ee
with
\bea
Y_{\pm} & = & (E_{q_1} + E_{q_2} + E_{q_1 \pm q_2})^2 \times (-
E_{q_1} + E_{q_2} + E_{q_1 \pm q_2})^2 \nonumber \\
& \times & (E_{q_1} - E_{q_2} + E_{q_1 \pm q_2})^2 \times
(E_{q_1} + E_{q_2} - E_{q_1 \pm q_2})^2
\eea
The effective mass $\overline{m}^2$ is a solution of the 2 loop
gap equation :
\bea
\overline{m}^2 & = & m^2 + \frac{\lambda \varphi^2}{2} +
\frac{\lambda}{2} \left[ \frac{\overline{m}^2}{16\pi^2} \left(
\ln \frac{\overline{m}^2}{\overline{\mu}^2} - 1 \right) + P_T
(\overline{m}^2)\right. \nonumber \\
& - & \frac{\lambda^2 \varphi^2}{6} \left. \left( \frac{\partial
G_0(\overline{m}^2)}{\partial \overline{m}^2} + \frac{\partial
G_1(\overline{m}^2,T)} {\partial \overline{m}^2} +
\frac{\partial G_2 (\overline{m}^2,T)}{\partial \overline{m}^2}
\right) \right]\label{gap}
\eea

\section{Numerical results}

The results obtained in the previous section were used to calculate the
effective potential for $\lambda\phi^4$-theory at the 2-loop 2PPI level 
numerically. These calculations involved two steps.
The first one is the numerical integration of the double integrals in 
(~\ref{F1}) and (~\ref{F2}).  Both integrals should be integrated from $0$ 
tot $\infty$. However we found that is was sufficient to approximate
the infinite integration interval by the compact 
interval $[0,10^5]$.  The stability of the numerical results towards 
changes in this interval and changes in the parameters of the model was 
checked and found to be satisfactory.
A second difficulty arises from the need to solve the gap equation which is a transcendental 
equation.  Trying to calculate the effective 2PPI potential as a function 
of $\varphi$, we need to solve the gap equation (~\ref{gap}) which gives us
the variational mass $\overline{m}$ as a function of $\varphi$. This could be done 
by iteration, but this approach will demand a lot of computational 
ressources.  We opted for a different approach, which consists of 
determining $\varphi$ as a function of the mass, calculate 
$V(\varphi(\overline{m}^2),\overline{m}^2)$ and plot $V(\varphi)$ by varying $\overline{m}$.

The parameters in our Lagrangian and the renormalization scale $\overline{\mu}$ were chosen in 
such a way
that comparison with the work of Chiku [13], which also uses an effective mass to improve upon 
the 1PI loop expansion, would be easy. We refer to this paper for more
details about this choice of parameters.  We used the two parameter sets used 
by Chiku in [13]
and found that the results did not depend qualitatively on the particular set. The numerical 
results in this
section have been obtained with $\lambda = 10.0, m^2=-170$ and $\overline{\mu}^2=87.6$.

In figure 4 the 1-loop 2PPI effective potential $V(\varphi)$ is shown for 
temperatures in the 
neighbourhood of the critical temperature.  This figure clearly indicates 
a first order 
transition.  Attention should be payed to the fact that this first order 
transition may easily 
be mistaken for a second order one if looked at at a larger scale.  
Therefore, it is necessary 
to use a sufficiently fine grid of points around $\varphi_0$ where the 
potential reaches its 
minimum, and to look at temperatures which are sufficiently close to the 
critical temperature.  
The transition also manifests itself as a discontinuity in the 
evolution of $\varphi_0$ at the 
critical temperature where $\varphi_0$ jumps from about 0.4 to 0. 
(figure 5).

Including the setting sun diagram, the 2PPI effective potential is altered 
to exhibit a second order phase transition, as can be seen from figure 6.  
The second order 
nature of this transition was extensively checked in the neighbourhood of 
the critical 
temperature $T_c=24.421$.  It is also evident from the continuous descent 
of $\varphi_0$ 
towards zero, where the symmetry is restored (figure 7).  In figure 7 there 
seems to be a 
shoulder structure for which we have no physical explanation.  

All of the above mentioned results show qualitative agreement with the 
2-loop analysis of 
$\lambda\phi^4$ theory in optimised perturbation theory, as done by Chiku 
in [13]

For the critical exponent $\beta$, which is defined by
\[\varphi_0(T) \sim \left|\frac{T-T_c}{T_c}\right|^\beta\]
for temperatures close to the critical temperature, we found the value 
$\beta = 0.52$ when taking the temperature interval $[24,24.421]$, 
indicating consistency with the Landau mean-field theory which predicts 
$\beta = 0.5$.  As a 
check for this last claim, we also investigated the critical behaviour of the 
second derivative of the 
effective potential and found again  accordance with the Landau mean-field 
prediction.

Summarizing, in this paper we have analysed the $\lambda\phi^4$ theory at finite
temperature using the 2PPI expansion at the 2 loop order. We found that the inclusion
of the setting sun diagram which is the only two loop 2PPI diagram, changes the phase
transition from first order at one loop 2PPI to second order at two loop 2PPI. The critical
exponents we found were mean field. We do not believe that adding more terms of the 2PPI 
expansion will improve the critical exponents. We expect that renormalization group resummation
of the 2PPI expansion [14] is necessary to breng the critical exponents in closer agreement
with experiment.

\newpage

\def\appendix{\par
\setcounter{section}{0}
\setcounter{subsection}{0}
\renewcommand{\theequation}{\Alph{section}\arabic{equation}}
\setcounter{equation}{0}}

\appendix
\section*{Appendix}

In this appendix, we will calculate and renormalize the setting
sun diagram. Because this is a 2PPI diagram, only the 2PPI
counterterms have to be taken into account. The 2PPR
counterterms are used to renormalize 2PPR insertions in 2PPI
diagrams. This is a different approach than ordinary 1PI
renormalization where all the 2loop diagrams (two bubble and
setting sun graphs) have to be considered together and all one
loop counterterms have to be inserted [10]. In [2] we showed
that both approaches are equivalent and the purpose of this
appendix is mainly to check this explicitely at two loops. For those aspects
of the calculations which are not related to renormalization, we
will borrow heavily from [9] to which we refer for further
details. 

Using the Saclay method, we find the following expression for
the setting sun diagram :
\be
I_{ss} = \frac{\lambda^2}{6} [G_0(\overline{m}^2) +
G_1(\overline{m}^2,T) + G_2(\overline{m}^2,T)]
\ee
where $G_0$ is the temperature independent part
\bea
G_0(\overline{m}^2) & = & \int \frac{d^D q_1}{(2\pi)^D} \int
\frac{d^D q_1}{(2\pi)^D} \frac{1}{q^2_1 + \overline{m}^2}
\frac{1}{q^2_2 + \overline{m}^2} \frac{1}{(q_1 + q_2)^2 +
\overline{m}^2} \nonumber \\
& = & 3 \int d[q_1,q_2] S(E_{q_1},E_{q_2},E_{q_3})
\eea
with
\be
d[q_1,q_2] = \frac{d^{D-1}q_1}{(2\pi)^{D-1}}
\frac{d^{D-1}q_2}{(2\pi)^{D-1}} \frac{1}{8E_{q_1}E_{q_2}E_{q_3}}
\ee
\be
S(E_{q_1},E_{q_2},E_{q_3}) = \frac{2}{E_{q_1} + E_{q_2} +
E_{q_3}} 
\ee
and $q_3 = - (q_1 + q_2)$ and where $G_1$ and $G_2$ are
temperature dependent with one and two Bose-Einstein factors
respectively :
\be
G_1(\overline{m}^2,T) = 3 \int d[q_1,q_2]n_B(q_1) \left[
S(E_{q_1}, E_{q_2} E_{q_3}) + S(-E_{q_1},E_{q_2},E_{q_3})\right]
\ee
\bea
G_2(\overline{m}^2,T) & = & 3 \int
d[q_1,q_2]n_B(q_1)n_B(q_2)\left[ S(E_{q_1},E_{q_2},E_{q_3}) +
S(- E_{q_1},E_{q_2},E_{q_3}) \right. \nonumber \\
& + & \left. S(E_{q_1},- E_{q_2},E_{q_3}) - S(E_{q_1},E_{q_2},- 
E_{q_3})\right]
\eea
The setting sun diagram at T = 0 has been calculated in [11,12]
in $D = 4 - 2\epsilon$~:
\be
G_0(\overline{m}^2) = \mu^{4\epsilon}
\frac{(\overline{m}^2)^{1-2\epsilon}} {(4\pi)^{4-2\epsilon}} 
\Gamma(2\epsilon
-1) \left( \frac{3}{\epsilon} + 3 - 8.966523919\epsilon +
...\right) 
\ee
Doing the integral over the angle between $q_1$ and $q_2$,
$G_1(\overline{m}^2,T)$ can be separated in divergent and finite
parts [9] :
\bea
G_1(\overline{m}^2,T) & = & \frac{3}{(4\pi)^2}
I^{\epsilon}_{\beta} (\overline{m}^2) \frac{1}{\epsilon} +
\frac{3} {(4\pi)^2} I^0_{\beta} (\overline{m}^2)\left[ 2 + \ln
\frac{\overline{\mu}^2}{\overline{m}^2}\right] \nonumber \\
& + & \frac{3}{4(2\pi)^4} \int^{\infty}_0 dq_1 \frac{q_1
n_B(q_1)}{E_{q_1}} \int^{\infty}_0 \frac{dq_2}{E_{q_2}}\left[ q_2
\ln |\frac{X_+}{X_-}| - 4q_1\right]
\eea
with 
\be
X_{\pm} = (E_{q_1} + E_{q_2} + E_{q_1 \pm q_2})^2 \times (-
E_{q_1} + E_{q_2} + E_{q_1 \pm q_2})^2
\ee
and
\be
I^{\epsilon}_{\beta} (\overline{m}^2) = \mu^{4\epsilon} \int
\frac{dq^{3-2\epsilon}} {(2\pi)^{3-2\epsilon}} \frac{n_B(q)}
{E_q}
\ee
Because $G_2$ has a Bose-Einstein factor for each
momentumvariable, it is U.V. finite and after doing the angular
integration one finds :
\be
G_2(\overline{m}^2,T) = \frac{3}{4(2\pi)^4} \int^{\infty}_0 dq_1
\frac{q_1 n_B(q_1)}{E_{q_1}} \int^{\infty}_0 dq_2 \frac{q_2 n_B
(q_2)}{E_{q_2}} \ln |\frac{Y_+}{Y_-}|
\ee
with
\bea
Y_{\pm} & = & \left( E_{q_1} + E_{q_2} + E_{q_1 \pm
q_2}\right)^2 \times \left( - E_{q_1} + E_{q_2} + E_{q_1 \pm q_2}\right)^2
\nonumber \\
& \times & \left( E_{q_1} - E_{q_2} + E_{q_1 \pm q_2}\right)^2
\times \left( E_{q_1} + E_{q_2} - E_{q_1 \pm q_2}\right)^2
\eea

We now investigate the renormalization of the setting sun
diagram at finite temperature in some more detail. Since the
counterterm Lagrangian is
\be
\delta {\cal L} = \frac{1}{2} \delta Z(\partial_{\mu} \phi)^2 +
\frac{1}{2} \delta Z_2 m^2 \phi^2 + \frac{1}{4!} \delta Z_4
\lambda \phi^4
\ee
we find from
\bea
\parbox{20pt}{\begin{picture}(50,20)
\GCirc(10,10){5}{1}
%\Vertex(5,10){1.5} 
%\Vertex(15,10){1.5}
\Line(1,5)(5,10)
\Line(1,15)(5,10)
\Line(15,10)(19,15)
\Line(15,10)(19,5)\end{picture}}
+ \mbox{crossing} + \parbox{20pt}{\begin{picture}(20,20)
\Line(3,3)(10,10)
\Line(17,3)(10,10)\Line(17,17)(10,10)\Line(3,17)(10,10)\Text(10,10)[c] 
{$\otimes$} 
\end{picture}} & = & \mbox{finite} 
\nonumber \\
\parbox{20pt}{\begin{picture}(20,20)
\Oval(10,15)(5,4)(0) 
\Line(0,10)(20,10)
\end{picture}}
+ \parbox{20pt}{\begin{picture}(20,20)
\Line(0,10)(6.5,10) \Text(10,10)[c]{$\otimes$} \Line(13.5,10)(20,10)
\end{picture}} & = & \mbox{finite}
\eea
that at one loop
\be
\delta Z = 0 , \delta Z_2 = \frac{1}{32\pi^2} \lambda
\frac{1}{\epsilon} , \delta Z_4 = \frac{3}{32\pi^2} \lambda
\frac{1}{\epsilon} 
\ee
In [2] it was shown that
\be
\delta Z^{2PPR}_4 = \delta Z_2
\ee
so that
\be
\delta Z^{2PPI}_4 = \delta Z_4 - \delta Z_2 = \frac{1}{16\pi^2}
\lambda \frac{1}{\epsilon}
\ee
To renormalize the setting sun diagram which is 2PPI, we only
have to include 2PPI counterterms
\be
\parbox{20pt}{\begin{picture}(20,20)
\GCirc(10,10){7}{1}
%\Vertex(3,10){0.5}
%\Vertex(17,10){0.5}
\Line(0,10)(20,10)\end{picture}}
+ \parbox{20pt}{\begin{picture}(20,20)
\Oval(10,15)(5,4)(0) 
\Line(0,10)(6.5,10) \Text(10,10)[c]{$\otimes$} \Line(13.5,10)(20,10)
\Text(10,2)[c]{\tiny{2PPI}}
\end{picture}}
+ \parbox{20pt}{\begin{picture}(20,20)
\Line(0,10)(6.5,10) \Text(10,10)[c]{$\otimes$} \Line(13.5,10)(20,10)
\Text(10,2)[c]{\tiny{2PPI}}
\end{picture}} = \mbox{finite}
\ee
This equation determines the two loop 2PPI part of mass renormalization.
When using the 2PPI counterterm for mass renormalisation of 2PPI
diagrams, we should keep in mind to replace the ordinary mass
$m^2$ with the effective mass $\overline{m}^2$. This is correct
to all orders as was shown in [2]. Coupling constant
renormalization gives :
\bea
\parbox{20pt}{\begin{picture}(20,20)
\Oval(10,15)(5,4)(0) 
\Line(0,10)(6.5,10) \Text(10,10)[c]{$\otimes$} \Line(13.5,10)(20,10)
\Text(10,2)[c]{\tiny{2PPI}}
\end{picture}} & = & - \frac{\lambda}{2} \mu^{2\epsilon} \delta
Z^{2PPI}_4 \sum \int \frac{1}{q^2 + \overline{m}^2} \nonumber \\
& = & - \frac{1}{2} \frac{\lambda^2}{(4\pi)^2}
\frac{1}{\epsilon} \left( I_0(\overline{m}^2) +
I^{\epsilon}_{\beta} (\overline{m}^2)\right)
\eea
with
\bea
I_0(\overline{m}^2) & = & \frac{\overline{m}^2}{(4\pi)^2} \left(
\frac{4\pi \mu^2}{\overline{m}^2} \right)^{\epsilon} \Gamma
(\epsilon - 1) \nonumber \\
& = & \frac{\overline{m}^2}{(4\pi)^2} \left[ - \frac{1}{\epsilon}
+ \ln \frac{\overline{m}^2}{\overline{\mu}^2} - 1 +
0(\epsilon)\right] 
\eea
From (A.1) and (A.7), we have
\bea
\parbox{20pt}{\begin{picture}(20,20)
\GCirc(10,10){7}{1}
%\Vertex(3,10){0.5}
%\Vertex(17,10){0.5}
\Line(0,10)(20,10)
\Text(10,0)[c]{\tiny{$T = 0$}} \end{picture}} = - \frac{\lambda^2}{4}
\frac{\overline{m}^2}{(4\pi)^4} \left[ \frac{1}{\epsilon^2} +
\frac{3}{\epsilon} - \frac{2}{\epsilon} \ln \left(
\frac{\overline{m}^2}{\overline{\mu}^2}\right) \right] + \mbox{finite}
\eea
As a check on 2PPI renormalization, we find that the
$\frac{1}{\epsilon} \ln \overline{m}^2/\overline{\mu}^2$ term
coming from overlapping divergences gets nicely cancelled and
that at two loops
\be
\delta Z^{2PPI}_2 = \frac{\lambda^2}{4} \frac{1}{(4\pi)^4}
\left( \frac{1}{\epsilon^2} - \frac{1}{\epsilon}\right)
\ee
Analogously, we find from (A.1), (A.8) and (A.19) that the
temperature dependent divergent part gets exactly cancelled by
the 2PPI counterterm diagram. Doing the renormalization to
$0(\epsilon^0)$, one finds that the finite parts of $G_0$, $G_1$
and $G_2$ are given respectively by (18), (19) and (21) of
section 3.

\newpage
\begin{figure}
\centering
\includegraphics[width=10cm,height=10cm]{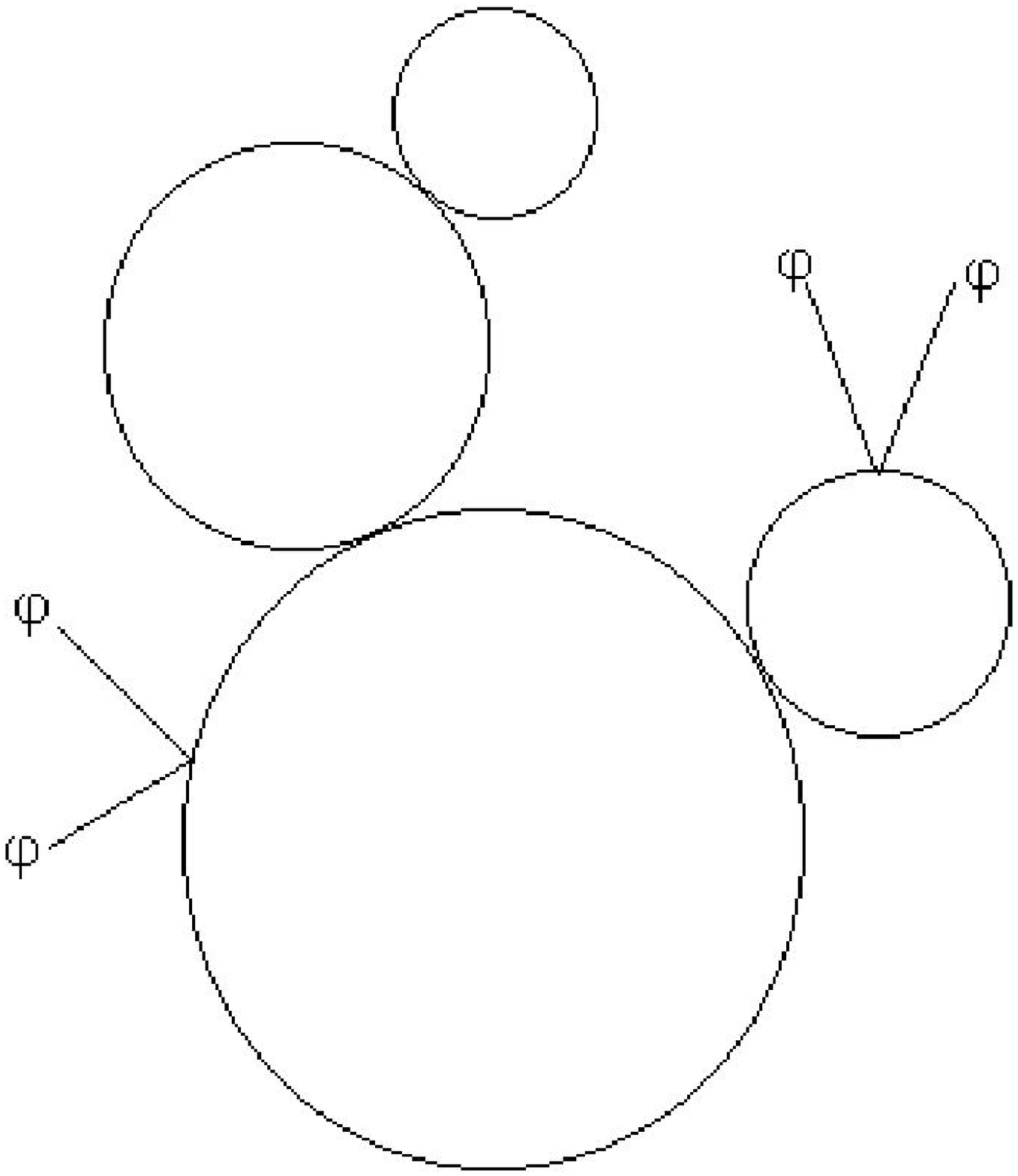}
\caption{Generic 2PPR diagram}
\end{figure}
\newpage
\begin{figure}
\centering
\includegraphics[width=10cm,height=12cm]{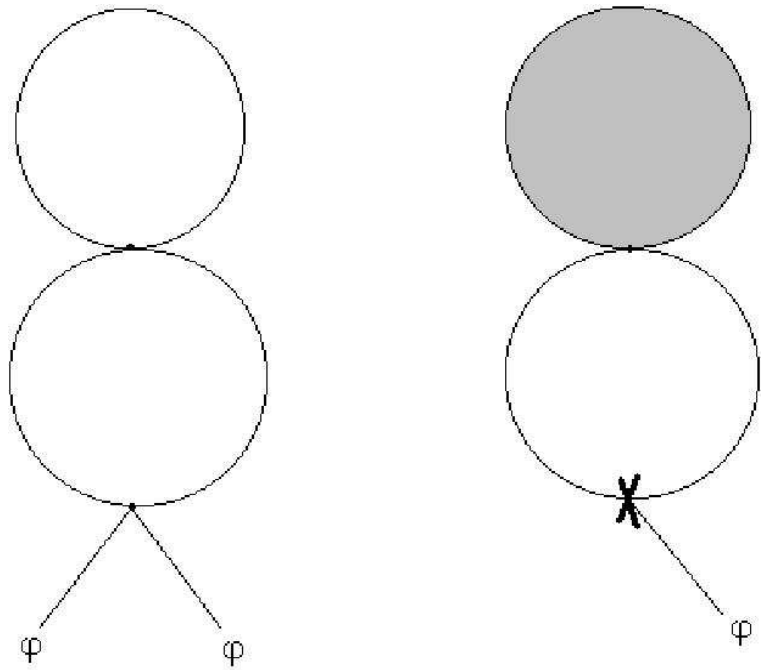}
\caption{2PPR part is shaded, 2PPI rest is earmarked}
\end{figure}
\newpage
\begin{figure}
\centering
\includegraphics[width=15cm,height=5cm]{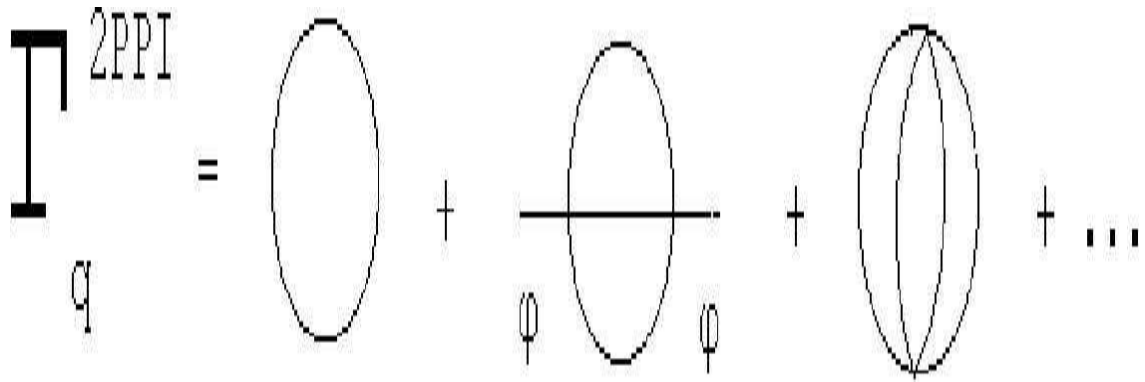}
\caption{First terms of 2PPI expansion}
\end{figure}
\newpage
\begin{figure}
\begin{center}
\includegraphics[scale = 1]{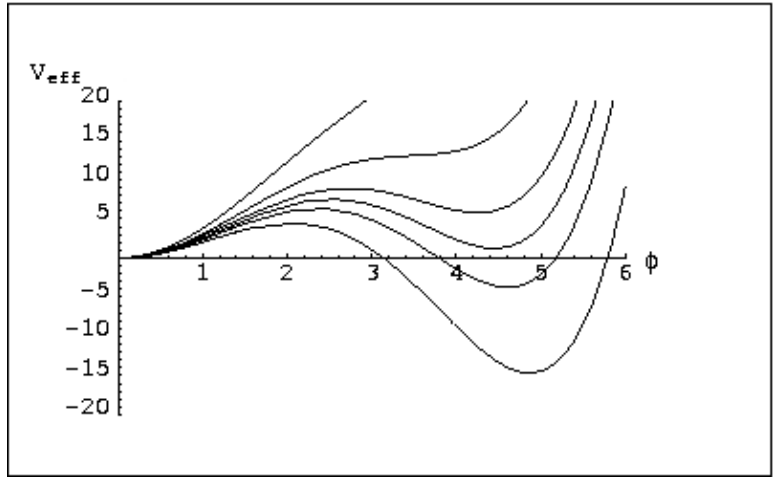}
\end{center}
\caption{{\footnotesize The 1-loop effective potential at 
T=21.5,21.6,21.65,21.7,21.8,22.  
At 1-loop we see a clear first order transition.}}
\label{fig4}
\end{figure}

\newpage

\begin{figure}
\begin{center}
\includegraphics[scale =1]{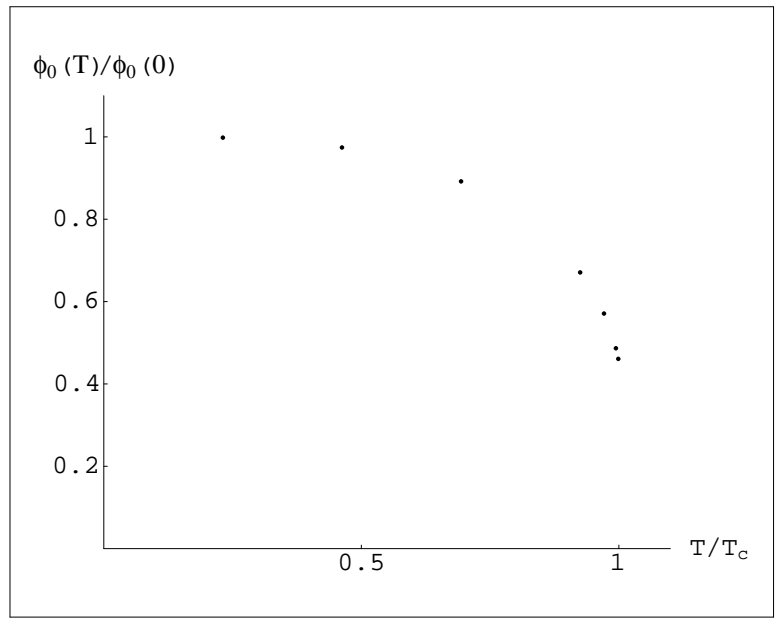}
\end{center}
\caption{{\footnotesize The 1-loop calculation of $\phi_0$ where the 
potential reaches it's 
minimum at different temperatures shows a discontinuous jump to zero at 
the critical 
temperature.}}
\label{fig5}
\end{figure}
\newpage

\begin{figure}
\begin{center}
\includegraphics[scale =1]{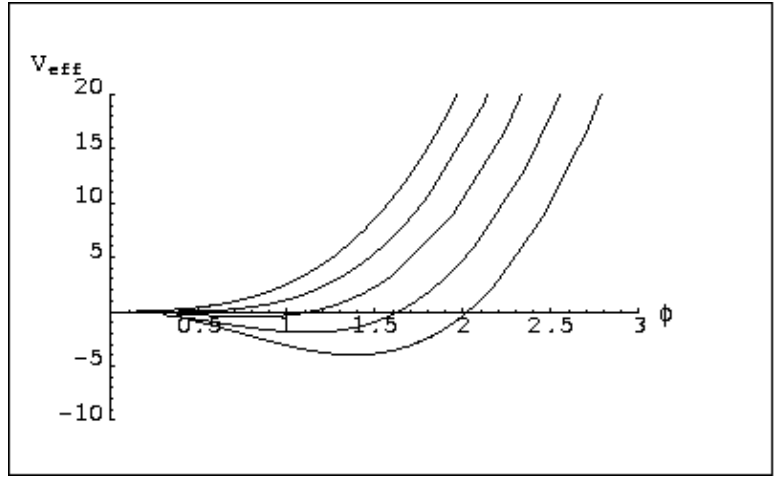}
\end{center}
\caption{{\footnotesize The 2-loop effective potential at T = 
23.8,24,24.2,24.4,24.6 .  
Including the SS-diagram changes the 1-loop first order transition to a 
second order one.}}
\label{fig6}
\end{figure}
\newpage
\begin{figure}
\begin{center}
\includegraphics[scale =1]{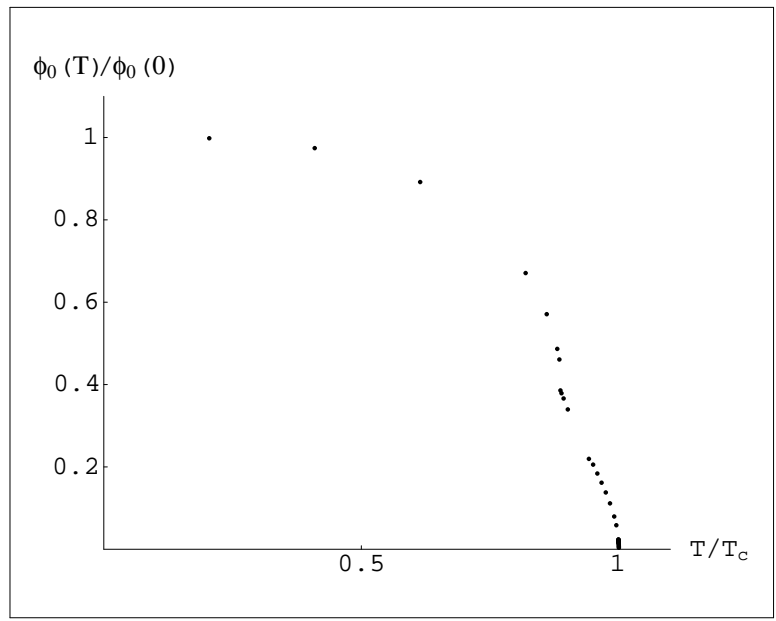}
\end{center}
\caption{{\footnotesize The 2-loop calculation of $\phi_0$ shows a 
continuous approach to zero 
at the critical temperature.}}
\label{fig7}
\end{figure}

\end{document}